\documentclass[12pt,superscriptaddress,onecolumn,preprintnumbers,balancelastpage]{article}

\usepackage{amssymb}
\usepackage{authblk}
\usepackage{booktabs}
\usepackage{epigraph} 
\usepackage{epstopdf}
\usepackage{color}

\definecolor{colorLink}{rgb}{0.6,0,0}
\definecolor{colorCite}{rgb}{0,.6,0}
\definecolor{colorURL}{rgb}{0,0.6,0.0}
\usepackage[pdftitle={Go rebco!},colorlinks=true,linktocpage=true,linkcolor=black,citecolor=colorCite,urlcolor=colorURL]{hyperref}
\usepackage{doi}

\usepackage[letterpaper, margin=1.2in]{geometry}
\usepackage[symbol]{footmisc}

\usepackage{graphicx}
\usepackage[numbers,sort&compress]{natbib}
\usepackage{url}

\newcommand{\U}[2]{\ensuremath{{#1}_{\mathrm{#2}}}}
\newcommand{\corc}{CORC\textsuperscript{\textregistered}}
\newcommand{\xin}{STAR\textsuperscript{\textregistered}}

\newcommand\snowmass{\begin{center}\rule[-0.2in]{\hsize}{0.01in}\\\rule{\hsize}{0.01in}\\
\vskip 0.1in Submitted to the  Proceedings of the US Community Study\\ 
on the Future of Particle Physics (Snowmass 2022)\\
\rule{\hsize}{0.01in}\\\rule[+0.2in]{\hsize}{0.01in} \end{center}}

\newcommand{\rb}{{\sc rebco}}

\title{{\rb} --- a silver bullet for our next high-field magnet and collider budget?}

\author[1]{\small Xiaorong Wang}
\affil[1]{\footnotesize \textit{Lawrence Berkeley National Laboratory, Berkeley, CA 94720, USA}}

\author[2]{\small Anis Ben Yahia}
\affil[2] {\footnotesize \textit{Brookhaven National Laboratory, Upton, NY 11973, USA}}

\author[3]{\small Ernesto Bosque}
\affil[3] {\footnotesize \textit{National High Magnetic Field Laboratory, Tallahassee, FL 32310, USA}}

\author[1] {\small Paolo Ferracin}
\author[1] {\small Stephen Gourlay}

\author[2] {\small Ramesh Gupta}
\author[1] {\small Hugh Higley}
\author[4] {\small Vadim Kashikhin}
\affil[4] {\footnotesize \textit{Fermi National Accelerator Laboratory, Batavia, IL 80510, USA}}
\author[2] {\small Mithlesh Kumar}
\author[4] {\small Vito Lombardo}

\author[1] {\small Maxim Marchevsky}
\author[1] {\small Reed Teyber}
\author[1,5] {\small Sofia Viarengo}
\affil[5] {\footnotesize \textit{Dipartimento Energia ``Galileo Ferraris'', Politecnico di Torino, Torino, Italy}}

\begin{document}
\pagestyle{plain}

\maketitle
\date
\snowmass


%
%




\pagebreak
%
%
\begin{abstract}
High-field superconducting magnets with a dipole field of 16 T and above enable future energy-frontier circular particle colliders. Although we believe these magnets can be built, none exists today. They can also be a showstopper for future high-energy machines due to a prohibitively high price tag based on the current conductor and magnet fabrication cost. The high-temperature superconducting \rb~coated conductor can address both the technical and cost issues, a silver bullet to lay both monsters to rest. The challenges and unknowns, however, can be too arduous to make the silver bullet. We lay out a potential road forward and suggest key action items. As a contribution from the accelerator community, we attempt to clarify for our theorist and experimenter colleagues a few aspects about the future high-field superconducting magnets. We hope to stimulate an effective plan for the 2023 P5 process that can lead to a cost-effective high-field magnet technology for future colliders and the exciting physics they can steward.
\end{abstract}


%
%
\setcounter{tocdepth}{1}
\tableofcontents

\pagebreak

\section{Executive summary}
\label{sec:summary}

\epigraph{The future cannot be predicted, but futures can be invented.}{\textit{Dennis Gabor}}

As an input to the Snowmass Community Study, we present this paper to our theorist and experimenter colleagues who use or consider using high-field superconducting magnets for their experiment and collider proposals. We hope to trigger more thoughts and discussions towards a sustainable and affordable future of high-energy colliders through the next P5 plan.

The paper concerns two issues: 1) the ultimate high field to meet the physics needs, and 2) the ultimate low cost to make our project affordable and funded! We think, from first principles, that \rb~can address both issues. First, \rb~material has a high irreversibility field with an upper limit of 110 T at 4.2 K and 100 T at 20 K. Such a high irreversibility can allow \rb~magnets to generate a dipole field of at least 20 T over a temperature range of 1.9 -- 20 K. Second, \rb~coated conductors have significant room for cost reduction due to the raw material cost. Given sufficient amount of production volume, the cost of \rb~conductor can reduce by an order of magnitude from today's \$100 kA$^{-1}$m$^{-1}$ to below \$10 kA$^{-1}$m$^{-1}$, approaching the cost of the commodity Nb-Ti conductor. Operating at elevated temperatures without liquid Helium can be another opportunity that \rb~can offer to further reduce the magnet and collider operation cost. 

Although \rb~shows great potential for addressing the high field and low cost needs, the challenges to realize the potential are significant for two reasons. First, we know little about \rb~magnet and conductor. Second, we do not have enough conductors to make magnets and to learn in a sufficiently fast pace. It is therefore important to radically focus on developing \rb~magnet technology within the next five years. We recommend the following:
\begin{enumerate}
\item Engage \rb~conductor vendors and couple the development of conductor and magnet technology. Use the magnet results as a critical feedback to the conductor development that, in turn, can help improve the magnet performance. 

\item Motivate and support multiple conductor vendors to meet the often challenging needs from magnet builders. Cultivate a healthy competition among vendors.
  
\item Significantly increase the funding for the U.S. Magnet Development Program to invest \$11 million a year for five years in \rb~development, including \$6 million a year on average for conductor development with multiple conductor vendors and \$5 million a year on average to make magnets. Leverage SBIR and allied programs to increase the funding.
  
\item When sufficient funding is available, fast-track the development with the goal to quickly progress on the maximum dipole field a \rb~magnet can generate. Aim for 10 T dipole field within three years and 15 T within five years. 

\item Encourage local and rapid development of a specific magnet concept at each lab and frequent exchange among labs for collective learning.

\item Support the growth of \rb-fusion symbiosis. Collaborate with the fusion magnet community and support the development of \rb~fusion magnet systems.
\end{enumerate}

\section{Bullet and beast}
\label{sec:introduction}

\epigraph{Of all the monsters that fill the nightmares of our folklore, none terrify more than werewolves, because they transform unexpectedly from the familiar into horrors. For these, one seeks bullets of silver that can magically lay them to rest.}{\textit{Frederic P. Brooks, Jr.}}

In a classic paper published in 1986~\cite{Brooks1987-tu}, Frederic Brooks Jr. described that a ``usually innocent and straightforward'' software project ``is capable of becoming a monster of missed schedules, blown budgets, and flawed products.'' He then asserted, through detailed analysis, that there is no silver bullet for the software-project-turned monsters. Here we attempt to treat the case of our next high-field magnets and collider budget. 

The monster is not unique to software project. As implied in the title of this paper, we think the innocent high-field magnets are capable of becoming beasts and causing troubles. By ``high-field magnets'', we mean magnets that can generate a dipole field of 16 T and above. High-field solenoid magnets are not discussed here, although they are critical to the proposed muon collider and dark matter experiments. The National High Magnetic Field Laboratory and other labs have been pioneering the development of high-field solenoid magnets. Recent achievements include the 32 T user magnet~\cite{Weijers14} and a record of 45.5 T dc field~\cite{Hahn2019-bd}.

We consider two troubles that a magnet monster can cause. First, fail to reach the ultimate capability to generate the desired dipole fields. A 16-T dipole magnet, although none exists today, is at the heart of future energy-frontier machines, enabling circular \textit{pp} colliders~\cite{fcc} and high-energy muon colliders~\cite{Al_ali2021-sp,Schulte2022}. Since the center-of-mass energy or luminosity scales linearly with dipole field, an ever higher dipole field is always in demand. 

The second trouble is the high cost of magnet fabrication and operation, even if the magnet can generate the desired dipole fields. The 16 T dipole magnets will cost about 39\% of the total cost to construct the proposed FCC-hh without prior implementation of FCC-ee~\cite{fcc}. The superconductor can dominate the cost of a high-field magnet, with Nb-Ti likely as the only exception~\cite{Vedrine2022}. Richter warned that ``without some transformational developments to reduce the cost of the machines of the future there is a danger that we will price ourselves out of the market.''~\cite{Richter2015-cm}

So can there be a silver bullet to lay rest both monsters? Breidenbach and Barletta first asked if \rb~can be a candidate~\cite{Breidenbach2020-fw}.

\section{Why {\rb}?}
\label{sec:promise}
\epigraph{It's time to see what I can do: to test the limits and break through.}{\textit{Princess Elsa of Arendelle}}

\rb, pronounced as [rebkou], is an abbreviation of REBa$_2$Cu$_3$O$_{7-\delta}$ where RE stands for rare-earth elements. The yttrium version, {\sc ybco}, has a high transition temperature, $\U{T}{c}$, of 93 K~\cite{Tinkham}. For magnet applications, \rb~refers to a composite material with a layered structure encapsulated by an electroplated Cu layer. We also call \rb~a ``coated conductor'' because the fabrication process essentially coat a thin \rb~superconducting layer onto a metallic substrate~\cite{Selva20121}. Table~\ref{tab:main_components} lists the major components in a \rb~coated conductor. The \rb~conductor is commercially available in a tape form with a width of a few mm and a thickness of 40 -- 100 $\mu$m. 

\begin{table}[!ht]
  \caption{Major components and their characteristic thickness in a typical commercial \rb~coated conductor relevant for high-field magnets.}
  \centering
  \begin{tabular}{rc}
    \toprule
    Component&  Characteristic thickness\\
             &  ($\mu$m)\\
    \midrule
    Cu &  10\\
    Ag &  1\\
    \rb &  1\\
    Ni-alloy substrate &  10\\
    \bottomrule
  \end{tabular}
  \label{tab:main_components}
\end{table}

Why we think \rb~coated conductors can be a silver bullet, albeit they have little silver (table~\ref{tab:main_components})? Let's start with the irreversibility field, $\U{H}{irr}$, at which the current-carrying capability of a superconductor vanishes~\cite{Larbalestier2001,Marken20121}. By definition, the maximum dipole field a superconducting magnet can generate is lower than the irreversibility field of the superconductor. \rb~has demonstrated a $\mu_0 \U{H}{irr}$ of 45 T at a temperature of 45 K~\cite{Miura2010-lu}, which doubles the irreversibility field of Nb$_3$Sn at 4.2 K~\cite{Larbalestier2001}. The $\U{H}{irr}$ increases with decreasing temperature~\cite{Tinkham}. Although we are not aware of the actual $\U{H}{irr}$ for \rb~below 45 K, we know its upper bound is 110 T at 4.2 K and 100 T at 20 K~\cite{Nakagawa1999-uu}. The high $\U{H}{irr}$ should allow \rb~magnets to generate a dipole field of at least 20 T over a temperature range of 1.9 -- 20 K. 

High $\U{T}{c}$ superconductors, including \rb, enjoy generally a high thermal stability margin compared to their low-temperature counterparts. The high stability margin is due to the larger difference between the critical and operation temperatures of a high $\U{T}{c}$ superconductor. The specific heat of magnet conductors and structural materials significantly increase as the temperature increases from 4.2 K. A \rb~magnet can absorb more heat without quenching~\cite{VanSciver2002}.

The implication is that \rb~magnets can provide sufficient magnetic fields to bend and focus particles for future hadron and muon colliders looking for magnets in the 16 -- 20 T range. The high stability margin and the capability to operate at a temperature above 4.2 K can be particularly suitable in applications with high thermal deposition to the magnet systems and fast-ramping magnets~\cite{Piekarz2019-am} that can be tricky for low-temperature superconducting magnets~\cite{Gupta2015-nl,Ogitsu2020-zz} .

The unique capability of \rb~is also recognized by the magnetic-confinement fusion community~\cite{Whyte2016,Maingi2019-op}. At least two private companies are now working on compact fusion devices, based on \rb~magnets, to demonstrate the fusion power generation.

\begin{figure}[!ht]
  \centering
  \includegraphics[width=0.5\columnwidth]{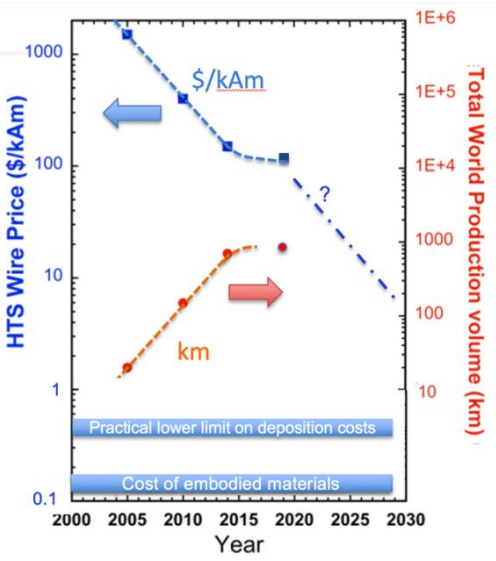}
  \caption{Price of \rb~tape and production volume over time. The cost of embodied materials and lower limit on materials deposition costs are also included. The dot-dash line is proposed by the authors of~\cite{Matias_LOI} to be feasible with the scale-up of tape manufacturing. Courtesy of Vladimir Matias.}
  \label{fig:price_price}
\end{figure}

We also think \rb~can ultimately address the increasing magnet and collider cost. The \rb~coated conductor has significant room for cost reduction. The raw material cost is low. Analysis from V. Matias and R. H. Hammond shows that given sufficient amount of production volume, the cost of \rb~conductor can reduce by an order of magnitude from today's \$100 kA$^{-1}$m$^{-1}$ to below \$10 kA$^{-1}$m$^{-1}$, approaching the cost of Nb-Ti conductor~\cite{Matias12,Marken20121,Matias_LOI,Uglietti2019-og}. Figure~\ref{fig:price_price} shows the price of \rb~tape and production volume over time, together with the raw material cost~\cite{Matias_LOI}. 

Another cost reduction opportunity is the operation of \rb~magnets at a temperature above 4.2 K, as recognized in the recent European Accelerator R\&D roadmap~\cite{Vedrine2022}. It is again enabled by the high $\U{H}{irr}$ of \rb~over a broad temperature range of 1.9 -- 40 K. Helium is a scarce resource with consistently climbing cost over the past few years with no clear trend to reverse. Operating at elevated temperatures also increases the cryogenic efficiency~\cite{Whyte2016}.

\section{Challenges}
\label{sec:challenges}
\epigraph{Superconductivity is absolutely the worst technology to use unless, of course, you have no other choice.}{\textit{David F. Sutter}}

Although from first principles, \rb~shows significant potential to address both the high field and low cost needs, the challenges to realize the potential are significant for two reasons. First, we know little about \rb~magnet and conductor. Second, we do not have enough conductors to make magnets and to learn in a sufficiently fast pace.

\subsection{We know little about \rb}
\label{sec:technical}

We know little about \rb~magnet and conductor, how they can be made, how they perform and how they fail, perhaps even more illiterate than we were with Nb-Ti in the 1970s. Two material properties of \rb~exacerbate the situation.

First, the ceramic \rb~material is brittle. Even though it is deposited on a strong Ni alloy substrate, the \rb~layer can crack, when subject to a tensile strain of around 0.6\% and higher, and permanently degrade its current-carrying capability. Bending or twisting the tapes that are necessary to wind coils or making multi-tape cable must not exceed this strain limit. For comparison, Nb-Ti can withstand a tensile strain around 1\% -- 2\%, depending on the applied field~\cite{Ekin1987-si}. \rb~coated conductors are also weak to withstand a tensile force applied transverse to its broad surface; a stress of several MPa can delaminate the tape and degrade the \rb~layer~\cite{Laan2007}. Epoxy impregnation can degrade \rb~conductor if the thermal contraction of the epoxy mismatches that of the conductor~\cite{Takematsu10}. Although no heat treatment is a relief, magnet builders now face a significant upfront risk when using \rb: we have every chance to degrade the brittle ceramic layer during magnet fabrication. 

Similar mechanical issues appear again during magnet operation when the Lorentz forces become excessive on the conductor. The maximum dipole field a \rb~magnet can generate will likely be determined by the mechanical limit of the conductor. One particular concern from the brittleness of \rb~is the potential performance degradation in conductor and magnet. Such a degradation has been observed in high-current \rb~fusion cable samples~\cite{Uglietti15,Bykovsky16b}.

Second, \rb~coated conductors are only available as a tape with an aspect ratio of at least 10 (\S~\ref{sec:promise}) whereas Nb-Ti is available as round wires. Round wires are isotropic in mechanical and electromagnetic behavior. They are easier to be assembled into multi-conductor cables or being wound into magnets with different shapes. The tape conductor can be bent, but only as a developable surface with limited flexibility, similar to bending a paper strip. This geometric constraint limits the potential magnet designs that one can work with \rb~conductors. One solution is a round-wire conductor form assembled from multiple tapes, such as \corc~\cite{Van_der_Laan2019-rz} and \xin~wires~\cite{Kar2019-lg}, both are actively pursued in the U.S. with a strong support from SBIR programs.

High-field dipole magnets require large-current conductors, operating at 10 -- 20 kA, to reduce the magnet inductance and to accelerate current ramping~\cite{Rossi12,Ferracin_LOI}. Various concepts exist for multi-tape \rb~cable for high-field magnets, such as twisted-tape stack~\cite{Takayasu12}, Roebel cable~\cite{Goldacker14}, \corc~wire~\cite{Van_der_Laan2019-rz} and \xin~wire~\cite{Kar2019-lg}.

CERN developed dipole magnets using a stack of tapes and Roebel cables~\cite{Van_Nugteren2018-rg,Rossi2021-qm}. The U.S. Magnet Development Program (MDP)~\cite{Gourlay16,Prestemon2020-hp} is developing dipole magnets using~\corc~wires with different magnet concepts~\cite{CC,Wang2018-df,Kashikhin2019-hq}. The current maximum dipole field achieved by a \rb~magnet is 5.4 T at 4.2 K in a racetrack magnet from the European EuCARD program and 4.5 T at 4.2 K in a Roebel-cable based magnet from the European EuCARD2 program~\cite{Rossi2021-qm}. 

There is a significant technology gap between where we are today and a \rb~high-field dipole magnet. Here we list six questions that need to be addressed for \rb~dipole magnet and conductor technology~\cite{Wang2019-pa}. 
\begin{enumerate}
\item How to make high-field accelerator magnets using multi-tape \rb~conductor?

  \rb~conductors are brittle and strain-sensitive, which can require specific magnet design and fabrication to minimize the strain-induced degradation. Magnet design and fabrication will help guide the conductor development: architecture, transport performance, bending radius, inter-tape contact and etc~\cite{Sumption_LOI_cable}. Impregnation and joint fabrication need to be addressed. High-current multi-tape cable development is a critical aspect for \rb~technology development~\cite{Ferracin_LOI,Uglietti2019-og}.

\item What is the maximum field a \rb~dipole magnet can achieve?

  What factors limit the maximum dipole field a \rb~magnet can generate and how can we address them? How to develop magnet structures to limit stress on conductors? What is the long-term performance of \rb~magnets under Lorentz loads? Will the performance under strong Lorentz forces degrade the conductor and magnet performance? 

\item How do \rb~magnets transition from superconducting to normal state and how can we detect the transition?

  The normal zone in \rb~magnets, once initiated, does not grow as fast as in a low-temperature superconducting magnet, challenging the quench detection and magnet protection against catastrophic damages. Innovative quench detection schemes will be required~\cite{Marchevsky2021-ja,Teyber2020-fz}.
  
\item What is the field quality of \rb~accelerator magnets?

  Field quality is what matters for particles. In addition to large magnetization in the conductors, its decay and impact on the accelerator operation needs to be understood and addressed~\cite{Sumption_LOI}.
  
\item What is the required performance for \rb~conductors to achieve the desired magnet performance?

  The conductor and magnet development are strongly coupled. We need to engage the allied material R\&D program and conductor manufacturers and help optimize the conductors by and for the magnet performance.
  
\item How to determine the performance of a long multi-tape \rb~conductor for predictable magnet performance?

  Accelerator magnets will require conductors with a unit piece length on the order of 100 m. The properties of \rb~conductors can vary along a long length. How do we characterize and improve its uniformity will be important for accelerator magnets~\cite{Hu17}.
\end{enumerate}

We note that not all these questions are created equal. The question is what questions we should focus on first.

\subsection{We have barely enough conductors to make mistakes}
\label{sec:cost_issues}

To know more and better about \rb, we have to build magnets, test and understand their performance. Ideally we make mistakes, learn and get better. Besides the many open technical questions, another significant obstacle to this essential learning process is a lack of \rb~conductors. Our experience so far with \rb~development is that we have barely enough conductors to make one magnet every several years. We cannot expect a baby to grow without sufficient food. As we try to get better and prepare to increase the dipole field for the next magnet, we find it more and more challenging to acquire enough \rb~conductors. Several factors contribute to the problem.

First, for dipole magnets, we use multi-tape cables based on single \rb~tapes. To make cable flexible enough for magnet use, we need thinner, narrower tapes with higher current than most of the commercial tapes available today. This is a significant technical challenge to tape manufacturers as it requires dedicated and potentially multi-year effort to meet our needs, although we are not yet picky in order to generate serious dipole fields above 10 T. The choice of rising to meet our challenges requires a vision and courage. It can be tough when everyone else is trying to sell tapes and cut cost. 

This leads to the second reason: \rb~dipole magnet development has only a limited number of tape vendor choices. Although there are a dozen or so companies worldwide offering \rb~tapes, only two or three of them today produce the tapes that cable manufacturers can use; and only one is in the U.S.

The last reason: \rb~conductor is so damn expensive relative to the available budget.

We should not be frustrated, however. What else do we expect for a new technology? Isn't this part of growing pain of \rb? What can we do to help tape manufacturers help us?

\subsection{Do we recognize the right problem?}
\label{sec:understand_the_problem}

There are a lot of questions and challenges about \rb. Some of them were given earlier. Of all these known problems, what is the most essential one that the U.S. MDP effort can best focus on in order to generate the strongest impact for the high-energy physics program? Only when we understand the problem, we can address it. For now, the problem is the maximum dipole field a \rb~magnet can generate and how quickly we can get there. 

We do not consider the conductor cost as a problem for us to directly address because, after all, the conductor cost won't be a problem if \rb~does not generate interesting dipole fields. The ultimate way to reduce cost is to create and keep enthusiastic customers by developing and demonstrating the unique capability of \rb~magnets.

We need to revisit the problems as we learn more about the \rb~and find more problems that are unknown today.

\section{There is a road}
\label{sec:path_forward}

\epigraph{A disciplined, consistent effort to develop, propagate, and exploit these innovations should indeed yield an order-of-magnitude improvement. There is no royal road, but there is a road.}{\textit{Frederic P. Brooks, Jr.}}

We will actually talk about two roads. One is to generate high fields with \rb~and the other is to reduce the \rb~cost. Let's first revisit a path, blazed by Wilson and his team, to achieve higher fields, followed by a parallel path to reduce cost that relies on a potential symbiosis between fusion magnets and \rb~conductors. We note road, path and approach are interchangeable here. 

\subsection{Wilson's path towards higher fields}
\label{sec:technology}


The curiosity about Wilson's approach is triggered by B. Richter's comment in 2015~\cite{Richter2015-cm}, ``I see no well-focused R\&D program looking to make the next generation of proton colliders more cost-effective. I do not understand why there is as yet no program underway to try to develop lower cost, high $\U{T}{c}$ superconducting magnets done on the scale of R. R. Wilson's efforts at Fermilab to successfully develop the first generation of commercially viable superconducting magnets that led to the Tevatron, HERA, and LHC.''

To revisit the magnet development in the Tevatron era can also be very instructive for \rb. We are as illiterate, if not more, about \rb~as we were about Nb-Ti conductor and magnet technology four decades ago.

In R. Wilson's own words, his approach ``..., largely Edisonian, was to build dozens and dozens of supermagnets, each only about one foot long but full scale in cross section. We built on our successes, tried to avoid repeating our failures, and accumulated experience; gradually the magnets improved until by now they are of quite adequate quality for an accelerator or a storage ring.''~\cite{Wilson1977-ve} With this and a paper from Sutter and Strauss~\cite{Sutter2000-nc}, we attempt to summarize the essential of Wilson's approach to magnet development.

\begin{itemize}
\item \textit{Set a clear goal}
  
  For Tevatron magnets, the goal is the highest possible magnetic field for the bending dipoles~\cite{Sutter2000-nc}. Tevatron needed 5 T and she had it. Our next colliders need 16 T, 20 T and perhaps higher. No matter what the number is, the highest possible dipole field should always be our goal for as long as we are still in the magnet business. We have a very simple indicator on how effective we are: a number with a unit of tesla.
  
\item \textit{Understand the problem}

  The Tevatron approach ``was that of industrial development rather than scientific research: How quickly could these magnets be reduced to practice for production?''~\cite{Sutter2000-nc} Tevatron needed a superconducting magnet that can quickly double the energy to do physics. It was a magnet that can be massively produced. Tevatron succeeded.

  For \rb, we think there are two problems, as discussed in \S~\ref{sec:understand_the_problem}. First, how quickly can we reach the ultimate dipole field with today's \rb~conductor? Second, what is the next right problem after we address the previous one? 
  
\item \textit{Rapidly build magnets}
  
  The Tevatron ``magnet development was Edisonian, that is, to rapidly build many models and test them to destruction.''~\cite{Wilson1977-ve,Sutter2000-nc}. To rapidly build magnets is the only effective way not only to make rapid progress but also to learn, especially for a new subject with significant unknowns. We cannot overemphasize this point for \rb. Fail quickly. Just like Nb-Ti, \rb~conductors are reacted, allowing a broad choices of tooling and options of construction; and the results can be available soon after we wind and cool down the magnets, allowing quick turn-around for fast learning. What is implied here is we have enough conductors to make mistake. 
  
  Some of us may frown upon the E-word, ``Gee, it doesn’t sound scientific.'' We certainly will leverage what we learned and the improved computer tools since the Tevatron era to carry out future research. The concern is that we risk missing the point if there is no working magnet, with higher fields, no matter what approach we take.
  
\item \textit{Accumulate experience}

  ``We built on our successes, tried to avoid repeating our failures, and accumulated experience.''~\cite{Wilson1977-ve} With the feedback from the magnet results, we understand what works and what needs improvement. This is especially important for \rb~conductor development that is largely unknown and strongly coupled with magnet behavior. By carefully design the experiments and control the variables we can learn what works and develop a working magnet and conductor technology. 
  
\item \textit{Adequate quality}

  Wilson and his team aimed for ``a magnet system that was \textit{adequate} for use in an accelerator.''~\cite{Sutter2000-nc} How much is adequate depends on our goal. For \rb, we focus on the dipole field. May we consider each magnet adequate if it generates a dipole field at least 2 T higher than its predecessor?

\end{itemize}

In short, take incremental but rapid steps to grow the \rb~magnet and conductor technology. Stick to the concept of ``minimum viable magnet'' with a focus on generating higher dipole fields. Understanding implications but avoiding premature optimization on important but secondary issues such as field quality. Building upon the experience and lessons learned from the previous steps, introduce and experiment few but new features in the next magnets towards a full set of magnet technology that can yield the ultimate dipole fields.

One necessary condition for the rapid magnet building and technology development is ``a readily available source of superconducting wire and cable.''~\cite{Sutter2000-nc} This is a big challenge for today's \rb~technology development (\S~\ref{sec:cost_issues}). Moreover, the optimal conductor form for high-field \rb~dipole magnets remains to be determined. To address this issue, we recommend the following:
\begin{itemize}
\item Engage \rb~conductor vendors and couple the development of conductor and magnet technology. Use the magnet results as a critical feedback to the conductor development that, in turn, can help improve the magnet performance. 
  
\item Motivate and support multiple conductor vendors to meet the often challenging needs from magnet builders. Cultivate a healthy competition among vendors.
  
\end{itemize}

The U.S. MDP is currently working on \rb~technology as one of the R\&D fronts~\cite{Prestemon2020-hp}. The budget, however, is inconsistent with MDP's broad R\&D scope. To maximize the progress of \rb~development, we recommend the following:
\begin{itemize}
\item Significantly increase the funding for the U.S. Magnet Development Program to invest \$11 million a year for five years in \rb~development, including \$6 million a year on average for conductor development with multiple conductor vendors and \$5 million a year on average to make magnets. Leverage SBIR and allied programs to increase the funding.
  
\item When sufficient funding is available, fast-track the development with the goal to quickly progress on the maximum dipole field a \rb~magnet can generate. Aim for 10 T dipole field within three years and 15 T within five years. 
  
\item Encourage local and rapid development of a specific magnet concept at each lab and frequent exchange among labs for collective learning.
\end{itemize}

\subsection{\rb-fusion symbiotic path towards lower cost}
\label{sec:cost}

Only with a market one can reduce product cost. Although Tevatron enabled today's MRI market for Nb-Ti conductors, future HEP machines will not likely to sustain a long-term market for superconducting materials or magnets~\cite{Uglietti2019-og}. We need to look outside HEP for solutions that can lead to affordable magnets~\cite{Gourlay18}. Fusion, specifically the magnetic confinement fusion that requires high-field magnets, can be another potential market for \rb, in addition to the usual suspect of MRI, particle therapy, and NMR market.

The latest fusion development has a strong push to develop smaller fusion devices to progress faster towards net-positive energy. The current fusion devices, including ITER, are usually huge devices developed as mega-projects that sometimes turn into the monster of ``missed schedules and blown budgets''. Progress has been historically slow, hence the joke of ``fusion is and will always be $x$ decades away,'' and hence the new push for smaller and sooner devices that are more nimble. Should not our future colliders also be smaller and sooner?

Since the fusion performance scales with $B^3$ or $B^4$ where $B$ is the on-axis magnetic flux density, one of the enabling technologies for the smaller and more powerful fusion device is, not surprisingly, \rb~magnets to generate a field beyond the reach of Nb$_3$Sn over a broad range of temperatures~\cite{Whyte2016}. This is exactly the same merit we are trying to exploit for high-field magnets for future colliders; a strong synergy between \rb~fusion magnets and HEP dipole magnets naturally appears. 

What could a \rb-fusion pair mean for our colliders? The answer relies on the possibility of them becoming our next symbiotic industries that can generate explosive growth for both. Indeed, after examining the symbiotic relationship between the microchip and computer industries, Isaacson pointed out in his book that ``There was a key lesson for innovation: Understand which industries are symbiotic so that you can capitalize on how they will spur each other on.''~\cite{Isaacson} Table~\ref{tab:symbiotic_industry} lists examples of symbiotic industries, including two from Isaacson.

\begin{table}[!ht]
  \caption{Existing symbiotic industries, including one from our own field that resulted directly from the legendary Tevatron. Will \rb~and fusion energy become a next symbiotic pair?}
  \centering
  \begin{tabular}{cc}
    \toprule
    Yin  & Yang \\
    \midrule
    Microchip  & Computer \cite{Isaacson}\\
    Oil & Auto \cite{Isaacson}\\
    Nb-Ti & Magnetic Resonance Imaging \cite{Strauss2011-te,Parizh2017-ag,Bruker}\\
    \rb & Magnetic confinement fusion ?\\
    \bottomrule
  \end{tabular}
  \label{tab:symbiotic_industry}
\end{table}

Historically, the symbiotic industries have caused the price of their products to fall rapidly and significantly. In addition to microchip and computer~\cite{Isaacson}, Nb-Ti and MRI offers another instructive example that resulted directly from the development of our own legendary Tevatron~\cite{Strauss2011-te}. Today, the principal market for Nb-Ti conductor is clinical MRI, with an approximately annual consumption of 4000 tons of Nb-Ti conductors~\cite{Bruker,Parizh2017-ag,Uglietti2019-og}. The Nb-Ti/MRI symbiosis explains why Nb-Ti becomes a commodity with a consistent and affordable price~\cite{Strauss2011-te}. An affordable conductor is possible, as long as there is a large market. 

The need for clean energy is pressing, and perhaps even more so, to some of us, than our need to understand how the universe works. Also the energy market is so universal and profound. No wonder the latest compact fusion development attracts significant public and private interests. For instance, Commonwealth Fusion Systems, a fusion startup in Boston, received in 2021 \$1.8 billion private investment to develop a compact tokamak system using \rb~magnets~\cite{CFS}. We speculate that magnetic confinement fusion, if successful, can create a sustainable market for \rb~conductors, leading to a reduced conductor cost. In fact, \rb~conductor vendors are already enthusiastically responding to the increasing needs from fusion~\cite{Molodyk2021-pm}.

Based on the strong synergies between \rb~fusion and dipole magnet technologies, we can and should proactively promote a strong symbiosis between the \rb and fusion industries. We should help fusion help us by supporting the development of \rb~fusion magnet technology. DOE Office of Fusion Energy Sciences and Office of High Energy Physics are jointly developing a 15-T large-aperture dipole magnet as part of a facility to test high-temperature superconducting fusion cables and dipole insert magnets~\cite{Vallone2021-dg}.

Although fusion and dipole magnets do not share the exactly same performance targets and characteristics, they deal with the same \rb~tapes and obey the same physics laws for magnet fabrication and operation~\cite{Uglietti2019-og,Hartwig2020-fw}. The flexible high-current \rb~cable that is required for small-aperture dipole magnets can be a useful building block for stellarator magnets featuring complex 3D shapes. The quench detection and protection technology that works for dipole magnets with a current density of more than 500 A mm$^{-2}$ can be useful to help protect fusion magnets with a lower current density. The multi-physics simulation and computational tools developed for \rb~magnets by either partner can be useful for the other. There are plenty of opportunities for us to compare notes, support and learn from each other.

To increase the chance of a significantly lower cost of \rb~conductor within the next decade, we recommend:
\begin{itemize}
\item Support the growth of \rb-fusion symbiosis. Collaborate with the fusion magnet community and support the development of \rb~fusion magnet systems.
\end{itemize}

\section{To make progress is not enough}
\label{sec:closing}

\epigraph{To make progress is not enough, for if the progress is not fast enough, something is going to overtake us.}{\textit{Leo Szilard}}

To conclude the paper and to continue the conversation with our theorist and experimenter colleagues, here is our last message: we need to make faster progress. Although tremendous progress has been made in \rb~magnet technology in the past few years, no \rb~magnet has yet generated a dipole field more than 5 T in a reasonable aperture and no compact fusion device has yet been built to generate net-positive energy. Until a serious dipole field beyond 16 T is demonstrated and fusion needs start driving down the cost of \rb~conductors, we doubt if \rb~can become a silver bullet. We definitely have a lot of progress to make.

And to make progress is not enough. It is not unusual for road maps of magnet technology development to span over a decade or longer. Magnet development takes time and time flies. But can we always assume we will have a decade or longer to develop the magnet technology? Although we won't need these magnets until decades from now, we need to demonstrate the feasibility of the technology as soon as we can in order to plan to actually build future colliders.


So, will \rb~become a silver bullet for our next high-field magnet and collider budget? We are skeptical, if not pessimistic, because of the painfully slow pace of \rb~technology development. We do know, however, that the answer is on our hands: it depends on how soon we can demonstrate the full potential of \rb~magnets.

We must have a sense of urgency to rapidly progress on affordable high-field magnets. Can we generate a dipole field of 10 T within 3 years and 15 T within 5 years using \rb? Can we make our colliders smaller and sooner using \rb? \textit {If progress is not fast enough, something is going to overtake us, our next high-field magnets, and colliders.}


\pagebreak
\section*{Acknowledgments}
The title of this paper is inspired by the presentation by M. Breidenbach and W. Barletta~\cite{Breidenbach2020-fw} and the paper by F. Brooks, Jr.~\cite{Brooks1987-tu}. We thank Vladimir Matias of iBeam Materials, Inc. for providing figure~\ref{fig:price_price}. The work at BNL was supported by Brookhaven Science Associates, LLC under Contract No. DE-SC0012704 with the U.S. Department of Energy. The work at FNAL is supported by Fermi Research Alliance, LLC under Contract No. DE-AC02-07CH11359 with the U.S. Department of Energy, Office of Science, Office of High Energy Physics. The work at NHMFL is supported by the National Science Foundation under Grant Number DMR-1644779, and by the State of Florida. The work at LBNL is supported by the U.S. Magnet Development Program through Director, Office of Science, Office of High Energy Physics and by Office of Fusion Energy Sciences of the US Department of Energy under Contract No.DEAC02-05CH11231.



\end{document}